\documentclass[oldversion]{aa}
\usepackage{graphicx}
\usepackage{natbib}
\usepackage{txfonts}
\bibpunct{(}{)}{;}{a}{}{,} % to follow the A&A style
\def\2237{Q2237+0305}

\begin{document}
   \title{The multiple quasar \2237 under a microlensing caustic\thanks{Based on
observations obtained with the Apache Point Observatory 3.5-meter telescope,
which is owned and operated by the Astrophysical Research Consortium.}}
   \titlerunning{A quasar under a microlensing caustic}
    \author{T. Anguita \inst{1}
	\and R. W. Schmidt\inst{1} 
	\and E. L. Turner\inst{2}
	\and J.  Wambsganss\inst{1}
	\and R. L. Webster\inst{3}
	\and K. A. Loomis\inst{4}
	\and D. Long\inst{4}
	\and R. McMillan\inst{4}
          }

   %\authorrunning{Anguita et al.}

   \offprints{T. Anguita, E-mail: tanguita@ari.uni-heidelberg.de}

   \institute{ Astronomisches Rechen-Institut, 
		Zentrum f\"{u}r Astronomie der Universit\"{a}t Heidelberg,
		 M\"{o}nchhofstrasse 12-14,
		 69120 Heidelberg, Germany.
               \and
               Princeton University Observatory,
               Peyton Hall,
               Princeton,
               NJ 08544,
               USA
% %	      \email{elt@astro.princeton.edu}
 	\and
 	School of Physics, 
	The University of Melbourne, 
	Parkville, 
	Victoria 3052, 
	Australia
  	\and
  	Apache Point Observatory, 2001 Apache Point Rd, Sunspot NM
                88349, USA}

   \date{\today}

   \abstract{We use the high magnification event seen in the 1999 OGLE campaign
light curve of image C of the quadruply imaged gravitational lens Q2237+0305 to
study the structure of the quasar engine. We have obtained g'- and r'-band
photometry at the Apache Point Observatory 3.5m telescope where we find that the
event has a smaller amplitude in the r'-band than in the g'- and OGLE V-bands.
By comparing the light curves with microlensing simulations  we obtain
constraints on the sizes of the quasar regions contributing to the g'- and
r'-band flux. Assuming that most of the surface mass density in the central
kiloparsec of the lensing galaxy is due to stars and by modeling the source with
a Gaussian profile, we obtain for the Gaussian width $1.20\times 10^{15}
\sqrt{{M}/{0.1M_\odot}}~cm \lesssim \sigma_{g'} \lesssim 7.96\times 10^{15}
\sqrt{{M}/{0.1M_\odot}}~cm$, where M is the mean microlensing mass, and a ratio
$\sigma_{r'}/\sigma_{g'}=1.25^{+0.45}_{-0.15}$. With the limits on the velocity of
the lensing galaxy from Gil-Merino et al. (2005) as our only prior, we obtain 
$0.60\times 10^{15} \sqrt{{M}/{0.1M_\odot}} ~ cm \lesssim \sigma_{g'} \lesssim 1.57 \times 10^{15} \sqrt{{M}/{0.1M_\odot}} ~ cm$ and a ratio 
$\sigma_{r'}/\sigma_{g'}=1.45^{+0.90}_{-0.25}$ (all values at 68 percent confidence). Additionally, from our microlensing
simulations we find that, during the chromatic microlensing event observed, the
continuum emitting region of the quasar crossed a caustic at $\geq72$ percent
confidence.}

\maketitle
%________________________________________________________________
%\maketitle

\section{Introduction}

	One of the best studied quasar lensing systems yet is \2237. This quasar
was discovered by the Center for Astrophysics (CfA) redshift survey
\cite[]{huchra85}. The system is made up of a barred spiral galaxy (z=0.0394)
and four images of a quasar at a redshift of z=1.695 that appear in a nearly
symmetric configuration around the core of the spiral galaxy at a distance of
approximately 1'' from the center.

	\2237 is an ideal system to study quasar microlensing. At the position
of the quasar images the optical depth due to microlensing by the stars is high
\cite[]{kayser86,kayser89,wambsganss90}. Furthermore the expected time-delay
between the images of this quasar is of the order of a day or less
\cite[]{schneider88, rix92,wambsganss94}. As microlensing events for this system
are of the order of months or years, it is easy to distinguish intrinsic
variability from microlensing. Another big advantage is the fact that the quasar
is $\sim$10 times farther from us than the lensing galaxy. This leads to a large
projected velocity of the stars in the source plane and thus a short timescale
of caustic events. Not long after its discovery in 1985, \cite{irwin89} observed
the first microlensing signature in this quasar.
	
The quasar has been monitored for almost two decades by different
surveys
\cite[e.g.,][]{corrigan91,pen93,ostensen96,alcalde02,schmidt02}. The
one that has the longest and best sampled data is that of the Optical
Gravitational Lensing Experiment (OGLE) team \cite[]{wozniak00,
udalski06}. They have monitored \2237 since 1997 delivering the most
complete light curves for this system in which many microlensing
events can be seen. These data have been used for various studies on
the system: \cite{wyithe00f,wyithe00g,wyithe00b,wyithe00d,wyithe00e};
\cite{yonehara01}; \cite{shalyapin02} and \cite{kochanek04} used this
data together with microlensing simulations not only to obtain limits
on the properties of the quasar such as size and transversal velocity,
but also on the mass of the microlensing objects.

The OGLE team has monitored this quasar with a single V band
filter. However, \cite{wambsganss91} showed that color variations
should be seen in quasar microlensing events and how they would
provide additional information on the background source. We know from
thermal accretion disk models that emission in the central regions of
a quasar should be bluer than in the outer regions
\cite[e.g.,][]{SS73}. As an
approximation we can assume the central engine of the quasar to be of
the order of 1000 Schwarzschild radii:
\begin{equation}
 R_{\rm ad}=1000\times r_s=2.8\times10^{16} \left(\frac{M_{\rm
BH}}{10^8M_{\odot}}\right){\rm cm}
\end{equation}
where $r_s$ is the Schwarzschild radius, $R_{\rm ad}$ is the radius of
the accretion disk and $M_{\rm BH}$ is the mass of the black hole. We
can compare this value with the Einstein Radius ($R_E$) - the length
scale in quasar microlensing - which can be given in the source plane
by \cite[]{SEF92}:

\begin{equation}
R_E= \sqrt{ \frac{4GM}{c^2} \frac{D_s D_{ls}}{D_l}} = 5.68 \times 10^{16}
\sqrt{\frac{M}{0.1 M_{\odot}}} {\rm cm} 
\label{eradius}
\end{equation}
where we have replaced the values of the distances $D_l$, $D_s$ and
$D_{ls}$ to the ones corresponding to \2237. 
The accretion disk scales are thus comparable to those of the Einstein rings of
stars in the lensing galaxy. This shows that differential magnification of
the different emission regions of the quasar can be observed.

	Some multi-band observations of this system have been carried out
\cite[]{corrigan91,vakulik97,vakulik04,koptelova05}, however most of them have
either large gaps between the observations, or no obvious microlensing event. 
	
	A very interesting event, that has been one of the bases for predictions
\cite[]{wyithe00g} and study \cite[]{yonehara01,shalyapin02, vakulik04}, is the
high magnification event of image C in the year 1999 as observed by OGLE.

In this paper we present data of this particular event 
obtained in two filters at Apache Point Observatory (APO).
We
analyze the event using both the OGLE light curve and the APO data
and study the implications for the size of the quasar emission region
using microlensing simulations.
We use a
flat cosmology with $\Omega_m=0.3$ and $H_0=70~km~s^{-1} Mpc^{-1}$.

\begin{figure}[]
	\centering
	\includegraphics[width=8cm,bb=0 0 732 598]{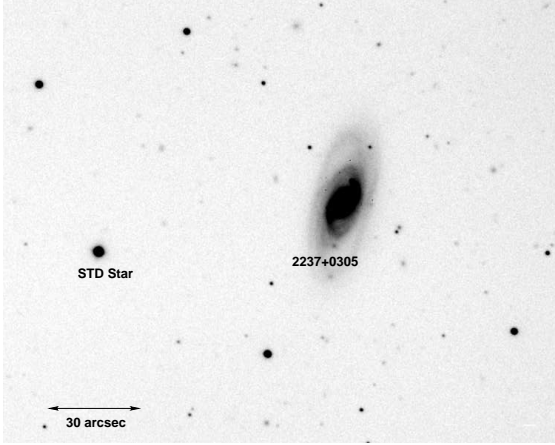}
	\caption{APO g' image of the quasar system \2237.}
	\label{full2237}
\end{figure}

\section{Apache Point observatory Data, Observations \& Data reduction}

For this study we use two different datasets. The first one is that of
the
OGLE\footnote{http://bulge.princeton.edu/$\sim$ogle/ogle2/huchra.html}
team light curves for this quasar \cite[]{wozniak00,udalski06} for
which we select the data between Julian days 2451290 (April 21, 1999)
and 2451539 (December 26, 1999), which comprises 83 data points. The
result is a dense light curve for the event we analyze. The second
data set is APO two-band monitoring data
for \2237. We selected the nights between May 17 1999 (Julian day:
24511316) and January 8 2000 (Julian day: 24511551) which also track
the event, comprising 8 nights with data points.

The APO data were taken with the 3.5 m telescope at Apache Point Observatory
using
the Seaver Prototype Imaging camera (SPIcam) in both the SDSS g' and
r' filters. The SPIcam has a 2048$\times$2048 pixels CCD and a minimum
pixel scale of 0.141 ''/pixel, however, for the observations we use,
the pixels were binned to 0.282 ''/pixel.

\subsection{Standard CCD Reduction}

	For the standard CCD reduction we use a combination of the astronomical
image reduction software package \textsc{iraf} and \textsc{idl} data language.
We apply bias correction, create one master flat-field image for each night and
each filter and calibrate the science frames with these. Cosmic rays are
eliminated from the science frames using median filtering outside the sources.

In order to remove the remaining cosmic rays and imperfections of the
CCD as well as to improve the signal to noise ratio of the images, we
combine the frames corresponding to the same night, excluding those
with asymmetric point spread functions (PSFs).
In the end we obtain a clean high quality image for each night in g'
and r' filters.

\subsection{Photometry}

The photometry for this system is made difficult because the quasar
images are seen through the core of the lensing galaxy. Our first
attempt, following the procedure as in \cite{wozniak00}, was to use an
image subtraction method \cite[]{alard98}. However our field of view
lacks the high number of bright stars required in order to run a
successful correlation and PSF characterization for the
procedure. This is why we use \textsc{galfit} \cite[]{peng02}
instead. \textsc{galfit} is a galaxy/point source fitting algorithm
that fits 2-D parameterized image components directly to the images.

	\subsubsection{\textsc{galfit} model}

	For the galaxy we use the model shown by \cite{schmidt96} (see also
\citealt{trott02}) from HST F7815LP band data, in which the lensing galaxy is
parameterized with a de Vaucouleurs profile for the bulge and an exponential
profile for the disk. The bulge's de Vaucouleurs profile is constrained to a
4.1'' (3.1 kpc) scale length, a 0.31 ellipticity and a 77$^{\circ}$ position
angle. For the exponential profile we use a 11.3'' (8.6 kpc) scale length, a 0.5
ellipticity and a 77$^{\circ}$ position angle. These profiles are constrained to
have the exact same central position within the image and a fixed magnitude
difference.

	The quasar images are parameterized as point sources. The relative
separation between them is fixed in the \textsc{galfit} code to the values
obtained by \cite{blanton98} from UV data and the separation between the group
of images and the center of the bulge is fixed to the values obtained with HST
F7815LP filter data (HST proposal ID 3799): $\Delta_{RA}=-0.075''$ and
$\Delta_{DEC}=0.937''$.

\begin{figure}
 \centering
 \includegraphics[width=8cm,bb=0 0 426 709]{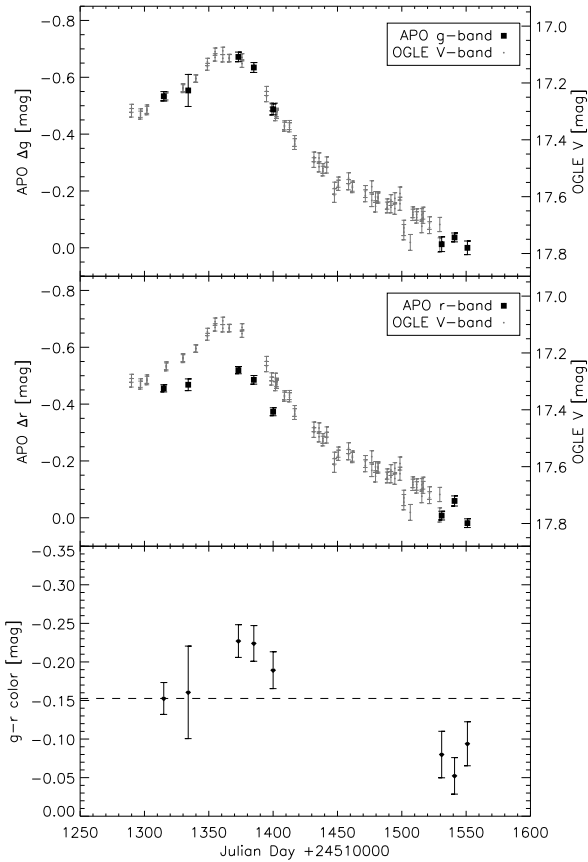}
 \caption{Top panel: Black squares show the APO g'-band light
 curve (relative to the faintest data point) of the high
 magnification event seen in image C of \2237\ in 1999. For
 comparison, in light gray, the OGLE V-band light curve sampling this
 event is shown. Middle panel: APO r'-band light curve for the
 event (black squares) and the OGLE V-band light curve plotted in
 light gray. Bottom panel: APO g'-r' color curve. The dashed line is
 shown to guide the eye.}
 \label{lightcur}
\end{figure}

The PSF we use to convolve the different parameterizations is taken
from the bright star (STD) shown in Fig. \ref{full2237}. After fitting
the different profiles, the brightness of each component is
measured. We run the \textsc{galfit} routine over each night in each
one of the two bands available. The residuals from this fitting
technique are low, usually of the order of 5 percent. In order
to obtain error bars on the magnitude measurements, we create 500
Monte-Carlo realizations of the observed images and determine the
measurement uncertainty from the scatter of the \textsc{galfit}
results. The g'- and r'-band light curves for image C are shown in
Fig. \ref{lightcur}. The measurements for images A and C are given in
Table \ref{tablightr}. No reliable photometry is obtained for images
B and D because they were too faint.

As shown in Fig. \ref{lightcur}, the APO g'-band light curve shows a
very similar behaviour to the OGLE V-band light curve, however, the
APO r'-band light curve shows a lower amplitude at the brightness
peak. Thus, the resultant g'-r' color curve, does not appear flat but
shows a chromatic variation during the microlensing event.
\begin{table*}[]
\renewcommand{\arraystretch}{1.0}
\centering
\begin{center}
\caption{\label{tablightr} APO light curves for quasar images A and
C. Magnitudes are shown relative to the value of the last data point
of image C in the g' filter.}

\begin{tabular}{c | c c c c | c c c c | c}

Julian Day&  \multicolumn{4}{c|}{A}   &  \multicolumn{4}{c|}{C} & \\
+2451000 &  \multicolumn{2}{c}{g'}   &  \multicolumn{2}{c|}{r'} & 
\multicolumn{2}{c}{g'}   &  \multicolumn{2}{c|}{r'}&  FWHM\\
      & $\Delta$ [mag]&err [mag]&$\Delta$ [mag]&err [mag] &$\Delta$ [mag]&err
[mag]&$\Delta$ [mag]&err [mag] &[arcsec]\\
\hline
        1315 & -0.822 & 0.016  & -0.675 &  0.012 &  -0.533 & 0.016   & -0.381 & 
0.013 & 1.6\\
        1334 & -0.826 & 0.056  & -0.680 &  0.021 & -0.554 & 0.056   & -0.393 & 
0.021 & 1.4\\
        1373 & -0.925 & 0.017  & -0.729 &  0.013 & -0.672 & 0.017   & -0.445 & 
0.013 & 1.2\\
        1385 & -0.971 & 0.017  & -0.782 &  0.015 & -0.635 & 0.017   & -0.411 & 
0.015 & 1.0\\
        1400 & -0.912 & 0.019  & -0.757 &  0.014  & -0.487 & 0.019 & -0.298 & 
0.014 & 1.1\\
        1531 & -1.148 & 0.024  & -0.887 &  0.015 & -0.013 & 0.026   &  0.067 & 
0.015 & 1.1\\
        1541 & -1.158 & 0.015  & -0.909 &  0.015  & -0.037 & 0.016 &  0.016 & 
0.018 & 1.1\\
        1551 & -1.137 & 0.018  & -0.916 &  0.014 &  0.000  & 0.024  &  0.094 & 
0.015 & 1.2\\
\end{tabular}
\end{center}
\end{table*}

\section{Microlensing Simulations}

	In order to compare the observed data with simulations we generate
source-plane magnification patterns for quasar images C and B using the inverse
ray shooting method \cite[]{wambsganss90,wambsganss99}. We assume identical
masses for all the microlenses distributed in the lens plane (the results do not
depend on the mass distribution of the microlenses \cite[]{lewis96}).
Approximately $10^{11}$ rays are shot, deflected in the lens plane and collected
in a 10,000 by 10,000 pixels (equivalent to 50 Einstein radii by 50 Einstein
radii) array in the source plane. The values of surface density $\kappa$ and
shear $\gamma$ are taken from \cite{schmidt98} (Table \ref{tabrob}). The
convergence $\kappa$ is separated into a compact matter distribution
$\kappa_{\star}$ and a smooth matter distribution $\kappa_c$. Even though we
explore different stellar fractions, as the quasar images are located within the
central kiloparsec of the lensing galaxy, we focus our work mainly on the
$\frac{\kappa_{\star}}{\kappa}=1$ stellar fraction.

\begin{table}[]
\renewcommand{\arraystretch}{1.0}
\centering
\begin{center}
\caption{\label{tabrob} Local macro-lensing parameters for quasar
images B and C from \cite{schmidt98}. $\kappa$ is the scaled surface
density, $\gamma$ is the shear, $\mu_{tot}$ is the total magnification.
}

\begin{tabular}{l l l l l l}
\hline
Image&  $\kappa$   & $\gamma$ & $\mu_{tot} $ \\
\hline      
B	&0.36 &0.42&4.29\\
C	&0.69&0.71&2.45\\
\hline
\end{tabular}

\end{center}

\end{table}

As shown in Sect. 1, in quasar microlensing the source size is an
important parameter because it is comparable with typical caustic
scales \cite[e.g,][]{kayser86}. In order to study a sample of
different source sizes, the raw magnification patterns are convolved
with a set of profiles. \cite{mortonson2005} showed that microlensing
fluctuations are relatively insensitive to the source shape, so it is
of little consequence whether we choose a standard model accretion
disk \cite[]{SS73} or a Gaussian profile. For simplicity we choose a
Gaussian profile for the surface brightness where the extent of the
source is described by the variance $\sigma$. The full width half maximum
(FWHM) is defined as 2.35$\sigma$.

We vary the FWHM of a Gaussian profile from 2 to 120 pixels (which corresponds to
0.01 $E_R$ and 0.60 $E_R$, respectively) for quasar image B and from 2 to 216 pixels (which corresponds to
0.01 $E_R$ and 1.08 $E_R$, respectively) for quasar image C (the additional magnification patterns for image C
are used for the color curve fitting explained in Section 5). For patterns below 12 pixels the step size is 2 pixels, and for patterns above 12 pixels the step size is 4 pixels. We find that this linear sampling gives more accurate results when interpolating compared to a
logarithmic sampling.

%convolved with Gaussian profiles with FWHMs greater than 12 pixels, FWHMs are increased in steps of 4 pixels and for patterns convolved with Gaussian profiles with FWHMs from 2 to 12 pixels, FWHMs are increased in steps of 2 pixels in order to supress resolution problems. 

For a specific convolved pattern we can now extract light curves. By defining a
track (path of the source in the source plane magnification pattern) with a
starting point, direction and velocity, we extract the pixel counts in the
positions of the pattern defined by this track using bi-linear interpolation.
The values are then scaled with the magnification values of Table \ref{tabrob}
for each image and converted into magnitudes.

\section{Light curve fitting}

\subsection{OGLE V light curve}

	Since quasars vary intrinsically, we need to separate this variation
from microlensing. We therefore study the difference between two light curves
extracted from microlensing patterns for different images to eliminate this
effect and allow for an additional magnitude difference (the time delay is
negligible, see Sect. 1).

	Using the V band OGLE light curves sampling this event in both images C
and B of the quasar \cite[]{wozniak00, udalski06} we construct the difference
light curve C-B. Simulated light curves for each image are extracted using
tracks over microlensing patterns created independently for each image and
convolved with a Gaussian profile with the same size. By repeated light curve
extraction with variation of the parameters, the best fit to these are obtained
from the comparison with the OGLE C-B difference light curve. What is actually
fitted are the parameters that define the track from which both light curves are
obtained. This track is constrained to have identical direction and velocity in
both the patterns C and B, but the starting point can be different among the
different patterns. It is important to remark that the direction is set to be the
same in both patterns, taking into consideration that the shear direction
between images C and D is approximately perpendicular \cite[]{witt94}.

	For the minimization method we use a Levenberg-Marquardt least squares
routine in \textsc{idl}\footnote{http://cow.physics.wisc.edu/$\sim$craigm/idl/}.
This routine is  a $\chi^2$-based minimization based upon MINPACK-1
\cite[]{more80}. As we are using 68 percent values for our errors, the $\chi^2$
is simply calculated as:
\begin{equation}
\chi^2= \sum_{i=0}^{DOF} \left( \frac{y_i-f(x_i)}{\sigma^{err}_i}\right)^2
\end{equation} 
	where DOF is the number of degrees of freedom, $y$ is the measured
value, $x$ is the independent variable model and $\sigma^{err}$ is the one sigma
error of the particular measurement. In the case in which we fit the OGLE data
for this event, the number of degrees of freedom is 75. $y_i$ are the
differences between the OGLE light curves for images C and B, $\sigma^{err}_i$
are the uncertainties measured by the OGLE team, $x_i$ are Julian days and
$f(x_i)$ are the difference between of the two light curves extracted from the
source plane magnification patterns for images C and B convolved with an
identical source profile. We also include a parameter which we call $m_0$ that
accounts for the magnitude offset between the images.

	As the size of each one of the convolved patterns is very large compared
to the time scale of the chosen event (of the order of several hundred times),
and considering the fact that it is relatively easy to find good fits because of
the huge parameter space that such large patterns give, we need to determine a
high number of tracks for each pattern. To obtain a statistical sample, we
search for 10,000 tracks for each considered source size. The first guesses for
each one of the tracks are distributed uniformly. However, during the fitting
process the starting points of the tracks are constrained to stay within a
square of 200 pixels sidelength around the random initial value in order to
force a sampling of the whole pattern.
	After this minimization is done we have a track library containing
3$\times10^5$ fitted tracks (10,000 tracks for 30 different source sizes with
best-fitting velocity and magnitude offset). 

\subsection{APO Color Curve}

Fig. \ref{lightcur} shows that the OGLE V-band and the APO g'-band light curves
are very similar. Since the microlensing effect depends strongly on the source
size, we assume, for simplicity, that the OGLE V-band continuum region is of
similar size as that of the APO's g'-band. Here we determine the ratio of the
APO r'-band region and the g'-/OGLE V-band regions that follows from the color
curve in Fig. \ref{lightcur}.

While the parameters that define the tracks are fixed by the procedure shown in
the previous section, the two parameters that influence the shape of the color
curve in this step are the ratio between the sizes of the two regions and the
intrinsic color g'-r' of the quasar ($c_0$). These are the only parameters
allowed to vary during this step. As before, we use the patterns convolved with Gaussian profiles up to 120 pixels FWHM
for image C to obtain the g'-band light curve. To obtain the r'-band light curve
we use the full set of patterns for image C. During the fitting process we
interpolate the magnitude values of the extracted light curves in order to
obtain continuous values for the source size ratio. After this second fitting
procedure, the $\chi^2$ values of the track library are updated: 
$\chi^2=\chi_{OGLE}^2+\chi_{APO}^2$.

\begin{figure}
 \centering
 \includegraphics[width=8cm,bb=0 0 736 942]{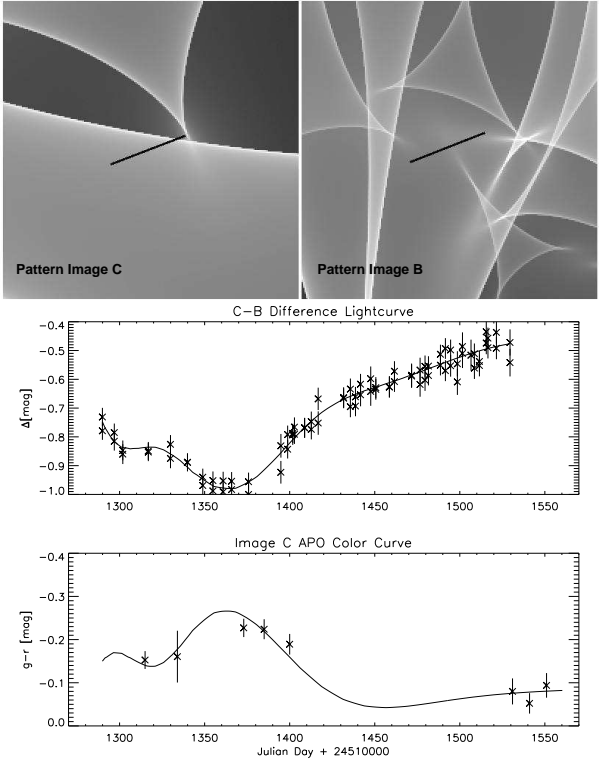}
 % 8221fig3.eps: 1179666x25 pixel, 300dpi, 9987.84x0.21 cm, bb=0 0 736 942
 \caption{Example track for a 0.05 $E_R$ (10 pixel) FWHM g'-band
source size and a source size ratio $\sigma_{r'}/\sigma_{g'}=1.48$. In the
top panels the fitted track is shown in a $1.0 E_R \times 1.0 E_R$
section of the microlensing patterns. In the middle panel we plot the
fitted difference light curve (solid line) and
the observed OGLE difference light curve. In the lower panel the best-fitting
color curve (solid line) and the APO g'-r' color curve are shown.}
 \label{fitting}
\end{figure}

\section{Considerations}

\subsection{Degeneracies}

	As source size and microlens mass are degenerate, we express the size
values in Einstein radii or scaled by a stellar mass as shown in equation
(\ref{eradius}). Another important degeneracy is the one between the size of the
source and the transversal velocity we fit. If we fix the path velocity of a
source and increase the size of it, its light curve will get broader.
Conversely, if we fix a source size and decrease the velocity, the light curve
will also become broader because it takes longer time for the source to cross
the magnification region.
	
\subsection{Velocity considerations}

The velocity one measures for a microlensing event is composed of
velocities of the observer, the lenses and the source.
The Earth's motion relative to the microwave background is likely to be
almost parallel to the direction of the quasar \2237 \cite[]{witt94}
so
the velocity of the observer
is negligible compared to the other velocity values.
We assume the physical
velocities of the lens and the source of the same order of magnitude,
thus, the main contribution to the total velocity is the velocity of
the lens due to the fact that the source velocity is weighted by
$\sim$3.5 percent compared to the lens velocity \citep[]{kayser86}.

The velocity of the lens has two components: one of the bulk velocity
of the galaxy $v_b$ and the velocity dispersion of the stars in the
galaxy (microlenses) $v_{\mu}$.
Since the time span of the event that we fit is
$\sim$200 days, the significance of individual stellar motions in
the fitting should be fairly low. Moreover, \cite{kundic93} and
\cite{wambsganss95} show that, in a statistical sense, the overall
effect of the individual stellar motions can be approximated by an
artificially increased velocity (a factor $\approx1.3$) of the source
across the pattern.

The scale we use for fitting the velocity is pixels per Julian
day. Using equation (\ref{eradius}) we can see that this turns into:
\begin{equation}
1 ~ \frac{pix}{jd}~=~0.005  ~\frac{E_R}{jd}=32870 \sqrt{\frac{M}{0.1 M_{\odot}}}
   \frac{km}{s}
\end{equation}
in the source plane.
Using the appropriate velocity scaling \citep{kayser86},
we find: 1 $\frac{pix}{jd }$ $\sim$
3000 $\frac{km}{s}$ in the lens plane.

\subsection{Size Considerations}

In a standard thin accretion disk model of a quasar the black hole is
surrounded by a thermally radiating accretion disk \cite[]{SS73}.
\citet{kochanek06} show that the radii $R_{\lambda}$ where
radiation with wavelength $\lambda$ is emitted scale as
$\lambda^{\frac{4}{3}}$. Thus:
\begin{equation}
f=\frac{R_{\lambda_1}}{R_{\lambda_2}}=\left(\frac{\lambda_1}{\lambda_2}
\right)^\frac{4}{3}.
\label{ratioeq}
\end{equation}
For the observed data from APO, the transmission curve for the SPIcam
g' and r' filters have central wavelengths of $\sim$4700\AA{} and $\sim$6400\AA{},
respectively. Using these values and equation (\ref{ratioeq}) we
obtain a predicted thin-disk ratio between the r'- and g'-band source
sizes of $\sim1.5$.
 
\section{Statistical Analysis}

 There are different ways to approach quasar microlensing studies. It can be by
analytical model fitting of the light curves \cite[e.g.,][]{yonehara01},
studying the structure function \cite[e.g.,][]{lewis96}, doing statistics of
parameter variations over time intervals \cite[e.g.,][]{schmidt98b,
wambsganss00, gilmerino05}, obtaining probability distributions with light curve
derivatives \cite[e.g.,][]{wyithe00a} or Bayesian analysis
\cite[e.g.,][]{kochanek04}. Similar to \cite{kochanek04}, we construct a
probability distribution for the quantities of interest (V/g'-band source size,
source size ratio and transversal velocity) from our track library.

Each one of our tracks is a fit to the light and color curves and
therefore has a $\chi^2$ value attached to it. We therefore infer
information from the whole ensemble of models: the track
library. Using the standard approach for ensemble analysis
\cite[e.g.,][]{sambridge99}, we assign a likelihood estimator to each
track $t_i$: $p(t_i) \propto \exp({\frac{\chi^2(t_i)}{2}})$.
Using this likelihood estimator the statistical weight for each track
can be written as:
\begin{equation}
 w(t_i)=\frac{p(t_i)}{n(t_i)}
\end{equation}
where $w(t_i)$, $p(t_i)$ and $n(t_i)$ are, respectively, the weight,
the likelihood obtained from the $\chi^2$ and the density parameter
for each track $t_i$. The density parameter describes the local track
density in a multidimensional grid in which each dimension corresponds
to a parameter of interest. The density is normalized so that the sum
of all the density parameters in the grid equals one. By summing over
the statistical weights, we can calculate probability distribution
histograms for all the parameters of interest.

\cite{gilmerino05} (from now on G-M05) found an upper limit of 625
$\frac{km}{s}$ (90 percent confidence) on the effective transverse velocity of
the lensing galaxy in \2237. They performed microlensing simulations assuming
microlenses with $M=0.1 M_{\odot}$ for three different source sizes yielding
similar results. We use their probability distribution for the velocity obtained
with the largest source size value (Fig. 5 in G-M05) in a further analysis step,
where we factor it into our own probability distributions by importance
sampling. In other words, we scale the probability of finding a track based on
the G-M05 prior, and then reobtain the best-fitting value for each parameter and
confidence regions.

\section{Results and Discussion}

\begin{table*}[]
\renewcommand{\arraystretch}{1.0}
\centering
\begin{center}
\caption{\label{tabresults} Results and confidence limits
for g'-band source size, ratio between r'- and g'-band source size and
transverse velocity (projected to the lens plane).}

\begin{tabular}{ c c c c c c c c c}
\hline

 \multicolumn{4}{c}{No Prior} & & \multicolumn{4}{c}{With G-M05 Prior} \\
  \multicolumn{2}{c}{Source Size $\sigma_{g'}$}  & Ratio  &  Velocity (*)&
&\multicolumn{2}{c}{Source Size $\sigma_{g'}$}   &Ratio &  Velocity\\[+1.ex]
$\times10^{-2}$  $R_E$& $ \sqrt{\frac{M}{0.1M_\odot}}$ &
$\frac{\sigma_{r'}}{\sigma_{g'}}$ & $ \sqrt{\frac{M}{0.1M_\odot}}$&
&$\times10^{-2}$  $R_E$ & $\sqrt{\frac{M}{0.1M_\odot}}$&
$\frac{\sigma_{r'}}{\sigma_{g'}}$ & $ \sqrt{\frac{M}{0.1M_\odot}}$  \\[+1.ex]
    & $\times10^{15}$[cm] & & [km/s]& & & $\times10^{15}$[cm] & & [km/s]  \\

\hline \\    
$8.10^{+5.90}_{-5.99}$ &$4.60^{+3.36}_{-3.40}$ &$1.25^{+0.45}_{-0.15}$ &$2930$& &
$2.34^{+0.43}_{-1.28}$ &$1.33^{+0.24}_{-0.73}$ & $1.45^{+0.90}_{-0.25}$
&$682^{+227}_{-379}$ \\[+1ex]
\hline
\end{tabular}

\end{center}
(*) The velocity obtained without the G-M05 prior is shown without
error bars as we do not obtain limits for it.
\end{table*}

\begin{figure*}[]
	\centering
	\includegraphics[width=18cm,bb=0 0 680 828]{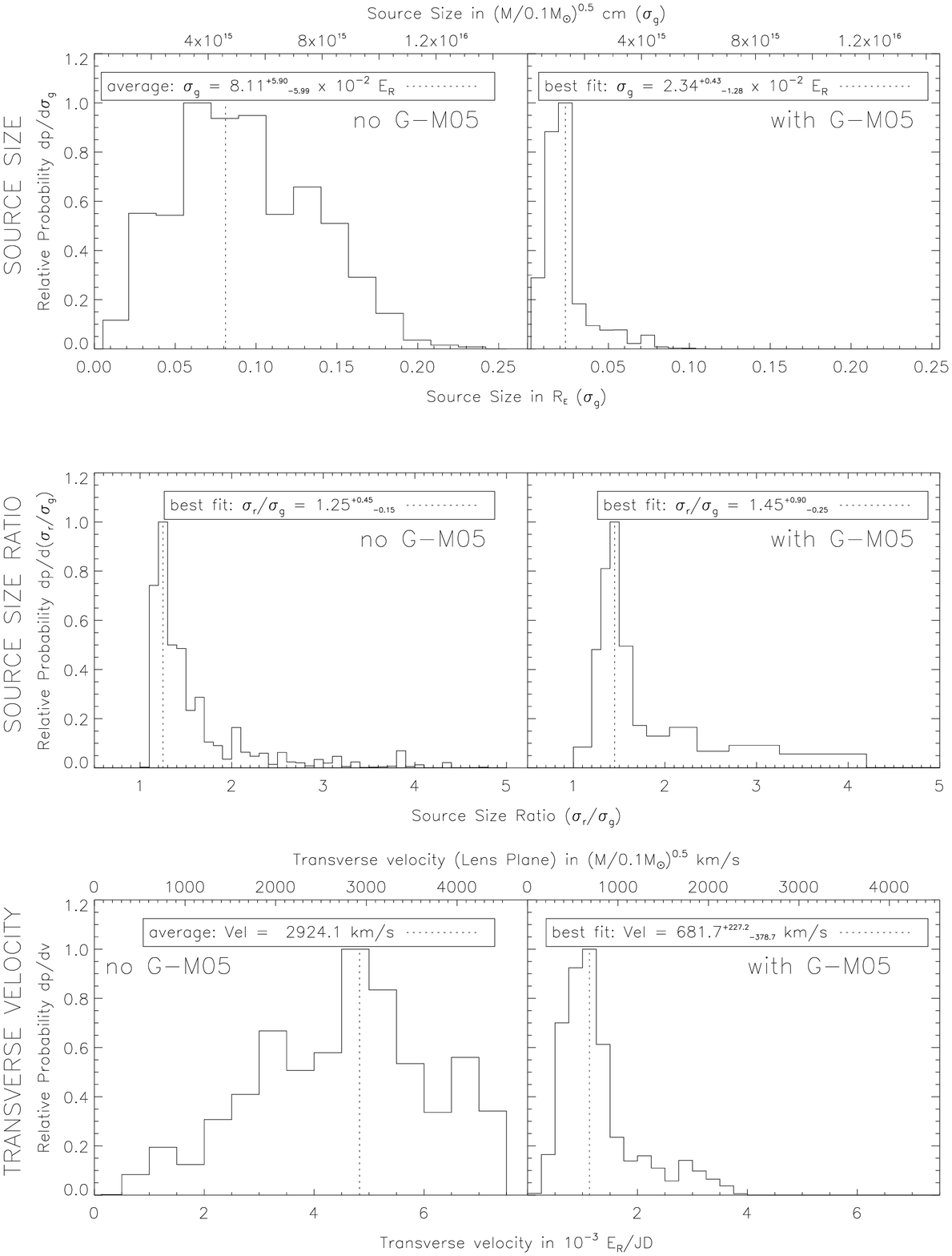}
	\caption{Probability histograms for g' or V source size, r'/g' source
size ratio and transverse velocity (projected to the lens plane), respectively.
The dotted line shows the best-fitting values (or average of the 68 percent
confidence region if no single best-fitting value is found). The error bars
describe the 68 percent confidence limits.}
	\label{histo}
\end{figure*}

	As described in the previous section, we obtain limits on the source
size and transverse velocity from two sets of probability distributions. The
first corresponds to the parameters obtained without using any prior, and a
second one in which we apply the G-M05 prior on the velocity of the lens galaxy.
For each probability distribution of a particular parameter we select the 68
percent confidence levels. This is done by making horizontal cuts starting from
the highest probability and screening down until 68 percent of the cumulative
probability is reached. The distribution, the best-fitting values (or the
average of the 68 percent confidence region where no single best-fitting
solution is found) and the 68 percent confidence limits for the relevant
parameters are shown in Fig. \ref{histo} and Table \ref{tabresults}. After applying the G-M05 prior only $\sim$1000 tracks carry 99 percent of the statistical weight. This makes the probability histogram for the source ratio sparsely populated for $\sigma_{r'}/\sigma_{g'}>2$. Therefore, we require at least 70 of these 1000 tracks to be part of each bin in the histogram.

	Without any velocity prior (see Fig. \ref{histo}), we have more fast
tracks than slow tracks. No limit on the transversal velocity can be determined.
This mostly is due to the degeneracy between source size and transverse velocity
(see Section 5.1). Using the G-M05 prior, we obtain a best-fitting velocity of
$682^{+227}_{-379} \sqrt{\frac{M}{0.1M_{\odot}}} km/s$ (projected into the lens
plane). Note, however, that the distribution shows a longer tail towards the
higher velocities when compared to the G-M05 probability distribution because
the steepness of the event fitted prefers fast tracks or small source sizes.

	 The average Gaussian width we obtain on the OGLE V/APO g' source size
without setting the velocity prior is $\sigma_{g'}=8.10^{+5.90}_{-5.99}\times
10^{-2} R_E$ or $4.60^{+3.36}_{-3.40}\times 10^{15}\sqrt{\frac{M}{0.1M_{\odot}}}
cm $. These limits agree with those obtained by \cite{yonehara01}
and \cite{kochanek04} (who used a logarithmic prior on the transversal
velocity). When we impose the velocity constraints of G-M05 we obtain an OGLE
V/APO g' source size with a Gaussian width of
$\sigma_{g'}=2.34^{+0.43}_{-1.28}\times 10^{-2} R_E$ which is equivalent to
$1.33^{+0.24}_{-0.73}\times10^{15} \sqrt{\frac{M}{0.1M_{\odot}}} cm$. This makes
the upper limit tighter by a factor of five compared to the previous result,
placing it just below the 68 percent confidence limit obtained by
\cite{kochanek04}, very close to their resolution limit. For $m_0$ we obtain $-1.75\lesssim m_0 \lesssim 0.25$ without and $-0.75\lesssim m_0 \lesssim 1.25$ with the G-M05 prior, respectively (see Section 4.1).

	The source size ratio between the r'- and g'-band emitting regions of
the quasar is  $\sigma_{r'}/\sigma_{g'}=1.25^{+0.45}_{-0.15}$ without, and  
$\sigma_{r'}/\sigma_{g'}=1.45^{+0.90}_{-0.25}$ with the G-M05 prior, respectively,
coupled with an intrinsic color parameter: $c_0=-0.18\pm0.05$ in both cases (see Section 4.2). As shown in Fig.
\ref{histo}, both source size ratio distributions are similar. The measurements are close to the
theoretical value (f$\sim$1.5) estimated in Section 5.3 from the \cite{SS73}
thermal profile.

For a long time there has been discussion on the nature of the image C
event in year 1999
\cite[e.g.,][]{wyithe00g,yonehara01,shalyapin02,kochanek04}. Using
different methods it was found that this event is likely to have been
produced by the source passing near a cusp. We investigate this for
the tracks in our library by computing the location of the caustics
for each of the magnification patterns using the analytical method by
\cite{witt90,witt91} \cite[combination of both ray shooting
simulations and analytical caustics are shown in][]{wambsganss92}.

	For each track we determine whether either the center, the
center$\pm\sigma_{g'}$ or the center$\pm\sigma_{r'}$ of the source touches a
caustic at any time. Our analysis agrees with the previous results when we use
the OGLE light curve to fit for this event. However, to reproduce our APO color
dataset, tracks that cross a caustic are favored (see Table \ref{tabcaucro}). We
find that a source with a radius as big as $\sigma_{r'}$ touched a caustic with
97 percent and 94 percent of confidence without and with the G-M05 prior,
respectively. Furthermore, the source, regardless of dimensions, crossed a
caustic with a 75 percent and 72 percent of confidence without and with the
G-M05 prior, respectively (i.e. the center of the source touched a caustic).

\begin{table}[]
\renewcommand{\arraystretch}{1.0}
\centering
\begin{center}
\caption{\label{tabcaucro} Probabilites of the image C high
magnification event being produced by the source crossing a caustic
and of the g' and r'-band region touching a caustic. The table shows
the obtained values for the probability both using and not using the
\cite{gilmerino05} velocity prior.}

\begin{tabular}{ c c c c c c}
\hline
\multicolumn{3}{c}{No Prior} & \multicolumn{3}{c}{With G-M05 Prior} \\
 cross  & g' &  r' &  cross  & g' &  r'  \\
\hline
0.75&0.92&0.97 &0.72&0.89 &0.94\\
\hline
\end{tabular}

\end{center}
\end{table}

	The probabilities of the source touching a caustic at different radii
are bigger in the case in which we do not use the G-M05 prior as seen in detail
in  Table \ref{tabcaucro}. This is due to the fact that a higher velocity makes
a track cover more distance through the pattern and thus is more likely to
encounter a caustic.

\section{Conclusions}

	We present two band (g' and r') APO data covering the high magnification
event seen in image C of the quadruple quasar \2237 in the year 1999. We find
that the amplitude of the brightness peak is more pronounced in the g'-band than
in the r'-band. This is also consistent with the observations by
\cite{vakulik04} (see also the recent analysis by \citealt{koptelova07}).

 By using this data together with the well known OGLE data
 \cite[]{wozniak00,udalski06} and combining it
 with microlensing simulations, we have been able to obtain limits on
 the size of regions of the quasar's central engine emitting in these
 bands: Gaussian width $\sigma_{g'}=4.60^{+3.36}_{-3.40}\times 10^{15}
 \sqrt{{M}/{0.1M_\odot}}$ cm and
 $\sigma_{r'}/\sigma_{g'}=1.25^{+0.45}_{-0.15}$.

	Because of the degeneracy between source size and transverse velocity we
use a prior on the velocity obtained from the work of \cite{gilmerino05} in
order to improve the results: Gaussian width
$\sigma_{g'}=1.33^{+0.24}_{-0.73}\times 10^{15} \sqrt{{M}/{0.1M_\odot}}$ cm and 
$\sigma_{r'}/\sigma_{g'}=1.45^{+0.90}_{-0.25}$.  Both values for the ratio between
the source sizes are close to the ratio obtained
for a face-on \cite{SS73} accretion disk (f$\sim$1.5). Recent studies
\cite[e.g.,][]{poindexter07,morgan07,pooley07} suggest microlensing yields a
slightly bigger value for the ratio than that obtained analytically with the
thin disk model (see Section 5.2), also in agreement with our results.

	 We also show that this event was probably produced by the source
directly interacting with a caustic as we obtain probabilities of 97 percent and
94 percent that the r'-band emitting region touches a caustic without and with
the G-M05 prior, respectively, and of 75 percent and 72 percent that the source
center (regardless of size) crossed a caustic without and with the G-M05 prior,
respectively.

\begin{acknowledgements}

TA would like to acknowledge support from the European Community's Sixth
Framework Marie Curie Research Training Network Programme, Contract No.
MRTN-CT-2004-505183 ``ANGLES" and the International Max
Planck Research School for Astronomy and Cosmic Physics at the
University of Heidelberg.

\end{acknowledgements}
\bibliographystyle{aa} % style aa.bst
\bibliography{8221}

\end{document}